# Danger is My Middle Name:
# Experimenting with SSL Vulnerabilities in Android Apps

Lucky Onwuzurike
Computer Science Department
University College London
lucky.onwuzurike.13@ucl.ac.uk

Emiliano De Cristofaro
Computer Science Department
University College London
e.decristofaro@ucl.ac.uk

## ABSTRACT

This paper presents a measurement study of information leakage and SSL vulnerabilities in popular Android apps. We perform static and dynamic analysis on 100 apps, downloaded at least 10M times, that request full network access. Our experiments show that, although prior work has drawn a lot of attention to SSL implementations on mobile platforms, several popular apps (32/100) accept all certificates and all hostnames, and four actually transmit sensitive data unencrypted. We set up an experimental testbed simulating man-in-the-middle attacks and find that many apps (up to 91% when the adversary has a certificate installed on the victim's device) are vulnerable, allowing the attacker to access sensitive information, including credentials, files, personal details, and credit card numbers. Finally, we provide a few recommendations to app developers and highlight several open research problems.

## 1. INTRODUCTION

Over the past few years, the proliferation of always-on, always-connected smartphones has skyrocketed. In 2013, 73% of mobile phone users regularly accessed the Internet via their mobile devices [7, 34] and this ratio is bound to increase as sales of smartphones and tablets keep growing [25, 26]. Naturally, as the number and the usage of always-on, always-connected smartphones increase, so does the amount of personal and sensitive information they transmit. Thus, it is crucial to secure traffic exchanged by these devices, especially considering that mobile users might connect to open Wi-Fi networks or even fake cell towers. However, SSL implementations in smartphone applications (in the rest of the paper simply referred to as *apps*) are actually more buggy and prone to vulnerabilities than browsers, including to man-in-the-middle (MiTM) attacks [21, 27]. Moreover, while browsers provide users with visual feedback that the communication is secured (via the lock symbol) and of certificate validation issues, apps do so less extensively and effectively [9].

Although prior work has presented means for detecting vulnerabilities [13, 33], it is unclear whether these are still prevalent in Android apps, and how so. We set to measure and analyze private information leakage and SSL vulnerabilities by building a sample of 100 *popular* Android apps requesting full network access (10% of all apps with at least 10M downloads, as per [2]). We examine them via static—decompiling using dex2jar [3]—and dynamic analysis, where we simulate three MiTM scenarios: (1) an advanced adversary that has its certificate installed on the user's device, (2) an SSL implementation accepting all certificates, and (3) an implementation not performing hostname verification correctly.

**Results.** We observe that almost all apps in our sample (93/100) include SSL, and most of them (78) use customized SSL code. Static analysis shows that half of the apps accept all certificates and half fail hostname verification. Dynamic analysis then reveals that 9 in 10 apps establish HTTPS connections under attack in scenario (1), while, in scenarios (2) and (3), about a quarter of them do so (23 and 29, respectively). Also, 4 apps—Deezer, Duolingo, Pic Collage, and 4shared—actually establish login sessions over HTTP. Finally, we report that only 3 apps provide relevant feedback indicating failure as a result of SSL certificate validation during an attack. As a consequence of these vulnerabilities, adversaries can access sensitive information, including credit card numbers, chat messages, contact list, files, and credentials. While we acknowledge that our work is similar in nature and, partly, in methodology to prior research [13, 21, 22, 27] (see Section 2), note that we do not only measure whether these are still prevalent, but also investigate and compare them via both static and dynamic analysis.

**Contributions.** In summary, we make the following contributions:

- We present a measurement study of Android apps security, confirming that SSL vulnerabilities are still prevalent even in very popular (10M+ downloads) apps.
- We investigate why static and dynamic analysis yield slightly different results.
- We provide recommendations for developers and highlight a few open research problems.

**Paper organization.** Next section reviews related work, then Section 3 presents some relevant background information. Section 4 introduces our methodology, while Section 5 discusses our measurement results, which we analyze in Section 6. Then, in Section 7 we discuss some recommendations, and conclude in Section 8. In the Appendix, we also provide a code snippet developers can be build from in order to implement SSL pinning as well as the complete list of analyzed apps.

## 2. RELATED WORK

Prior work looked at Android security from several perspectives: permissions [10, 11], privilege escalation [14, 18], detecting malicious apps [38, 40], information flow/taint tracking [20, 41], and details of SSL implementations [15, 21, 22, 27].

Georgiev et al. [27] study libraries and APIs used in SSL implementations of non-browser apps, and Brubaker et al. [13] introduce techniques for large-scale testing of certificate validation logic. Also, [15] and [33] introduce tools (respectively, MITHYS-App and SMV-Hunter) for automated vulnerability discovery.

A few studies [8, 9, 36, 39] analyze issues with SSL warnings in web browsers, but not many in the context of apps. In [21], Fahl et al. report that numerous apps do not display meaningful warning messages in presence of possible attacks, and, in [22], that 312 out of 599 apps display warnings, with 254 of them not being de-scriptive of the problem. MITHYSApp [15] also warns users about vulnerable apps and allows them to decide whether to continue the connection. Work in [21, 22] use static analysis to discover vulnerabilities in Android and iOS apps. Our work extends these by performing both static and dynamic analysis and investigate why they differ from each other. We also propose a fail-safe strategy that developers can use to implement SSL securely. Interestingly enough, although prior work, including [21,22,27], has highlighted important vulnerabilities in Android/iOS SSL implementations, 1-2 years later these are still prevalent even in very popular apps.

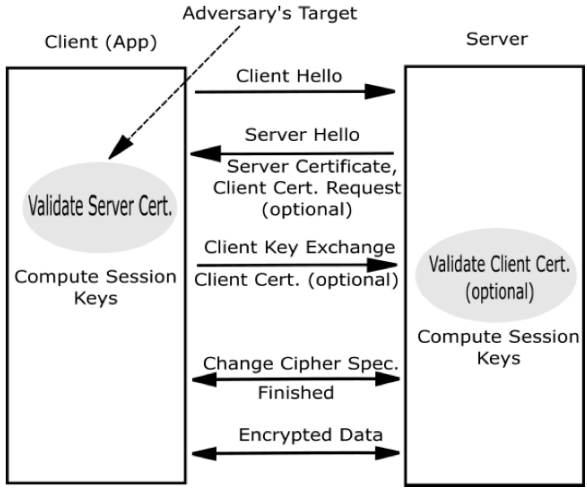

**Figure 1:** SSL Handshake.

Finally, Bates et al. [12] introduce CERTSHIM, a tool for certificate verification retrofitting that is dynamically hooked to SSL and data transport libraries, while Kranch and Bonneau [30] reveal that cookies can be leaked from domains that implement key pinning to malicious scripts in HTTP domains they load resources from.

## 3. BACKGROUND

In this section, we introduce some background information about SSL handshake, man-in-the-middle attacks, implementation of SSL on Android, and SSL Pinning.

**SSL Handshake.** SSL sessions always begin with a message exchange between the client and the server – the SSL handshake. Specifically, as depicted in Figure 1, the client initiates it by sending the *client hello* message and then verifies the identity of the server. Obviously, if the verification is not implemented correctly, an attacker could impersonate the server. The verification involves checking the Chain-of-Trust in server's certificate, hostname [29], and certificate validity [16,32]. It may also include verifying server's certificate serial number, key usage, etc.

**SSL MiTM Attacks.** In a man-in-the-middle (MiTM) attack, an attacker sits between two communicating parties, and intercepts, reads, modifies, and/or relays messages between them. In a MiTM attack over SSL, an attacker may also attempt to subvert the certificate verification in the SSL handshake, aiming to, e.g., read or modify encrypted traffic, and succeeding if:

- The client accepts all certificates as the server's,
- The client does not verify the identity of anyone claiming to be the server,
- The client accepts expired certificates,
- The server's certificate is forged by the adversary, or
- The client accepts all self-signed certificates.

**SSL Implementation in Android.** To guarantee end-to-end security, application developers use SSL over HTTP (HTTPS) sockets, relying on the android.net, android.webkit, java.net, javax.net, java.security, javax.security.cert, and org.apache.http packages of the Android SDK to create HTTP/HTTPS connection, administer, and verify certificates and keys. These packages provide developers with TrustManager and HostnameVerifier interfaces, which are used in the SSL certificate validation logic. TrustManager manages the certificates of all Certificate Authorities (CA) used in assessing the certificate's validity. Only root CAs trusted by Android are contained in the default TrustManager. The certificates are stored in a keystore which is then used to create the TrustManager. HostnameVerifier performs hostname verification whenever a URL's hostname does not match the hostname in the peer's identification credentials.

Developers can use default implementations of the TrustManager and HostnameVerifier interfaces provided by the Android SDK, or use customized ones. The latter usually involves developer-specified credentials or validation logic on TrustManager and/or HostnameVerifier. Typically, customization is done to support a certificate issued by a CA unknown to the OS, when using self-signed certificates, pinning peer credentials ("SSL pinning", see below), server configuration sending incomplete certificate chain, or the use of a single certificate for multiple hosts.

In the rest of the paper, we simply refer to the name of an interface or its implementation (e.g., TrustManager), whenever it is clear from the context.

**SSL Pinning** involves coupling a host's trusted credential (e.g., an X.509 certificate or a public key) to its identity. Any other credentials received other than that coupled will not be accepted as valid for the host, whether or not it was issued by a valid entity or an entity trusted by the OS. This is usually done when a developer knows, beforehand, the trusted credentials of the host.

## 4. METHODOLOGY

**Experimental testbed.** We use an LG Nexus 4 smartphone (running Android KitKat 4.4.4) with a Lenovo laptop (running Windows 8.1) acting as the Wi-Fi Access Point (AP). To capture traffic and simulate MiTM attacks, we use Wireshark [6] and Fiddler2 [4].

**Timeline.** Experiments presented in this paper were performed in August/September 2014, and then revisited in March/April 2015. All statistics refer to the first round of tests, however, whenever an app is no longer vulnerable, we report it in the text.

**Analyzed Apps.** Our first step is to build a set of 100 popular Android apps. We select 97 apps that request permission for full network access and that have been downloaded at least 10 million times. We choose apps from several different categories (including social networks, gaming, IM, e-commerce, etc) that access and process multiple kinds of sensitive information rather than those that only exchange login credentials. We also add 3 more apps that request full network access, even though they have less than 10M downloads: Barclays Mobile Banking, TextSecure, and Amazon Local. We choose Barclays Mobile and TextSecure motivated by the curiosity of analyzing at least one mobile banking and one secure chat app, and Amazon Local to verify the presence of any difference with the standard Amazon app.[1]

The total number of downloads for the 100 selected apps amounts to over 10 billion, according to statistics provided by the Google Play store. Table 1 summarizes the distribution of apps by number of downloads (at the time of our experiments). Note that, according to [2], there were about 1,000 apps with 10M+ downloads around the time of our experiments, thus, our 100-app set roughly corresponds to a 10% of all apps with 10M+ downloads. The complete list of apps we analyze is reported in the Appendix.

---
[1] Amazon and Amazon Local did implement SSL differently as for our first tests but both implement pinning as of our second tests.

| #Downloads (Millions) | #Apps |
|---|---|
| > 1,000 | 4 |
| 500 - 1000 | 6 |
| 100 - 500 | 31 |
| 50 - 100 | 19 |
| 10 - 50 | 37 |
| < 10 | 3 |

**Table 1:** Distribution of examined apps by number of downloads.

**Sensitive information.** Throughout our experiments, we classify as sensitive information traffic that includes: login credentials, device information (e.g., IMEI number), location data, chat and email messages, financial information, calendars, contact list, files, and so on. We exclude sensitive data transmitted by embedded ad libraries as this is out of the control of the app developer.

**Static Analysis.** We decompile apps using dex2jar [3] and JD-GUI [5] and search key terms such as `HttpsURLConnection`, `HostnameVerifier`, and `TrustManager`, which indicate the presence of SSL code. Then, we analyze the `TrustManager` and `HostnameVerifier` implementations used by the apps.

**Dynamic Analysis.** For the dynamic analysis, we proceed to use features of the apps that would request/need sensitive information and probe for vulnerabilities. We assume an adversary that has control over the Wi-Fi access point the victim connects to and simulate the three following possible MiTM attack scenarios:

**S1:** The adversary has his certificate installed on the user's device (to simulate this, we install a Fiddler certificate as root CA);

**S2:** The adversary presents an invalid, self-signed certificate;

**S3:** The adversary presents a certificate with a wrong Common Name (CN) and/or SubjectAltName, signed by a root CA.

**Ethics.** Note that we conducted all experiments in a controlled test environment and did not monitor or capture any user's traffic.

## 5. RESULTS

### 5.1 Static Analysis

Via static analysis, we find that 93/100 apps include SSL code, while the remaining 7 (AVG antivirus, Candy Crush, Clean Master, Google Earth, Play Music, Jobsearch and Voice Search) do not include SSL, even though they use tokens to access secure web pages using Google, Facebook, or Twitter user accounts on the device. 84% (78/93) of the apps with SSL implement their own `TrustManager` or `HostnameVerifier`, while the remaining 16% use Android's default `TrustManager` and a combination of the subclasses of the `HostnameVerifier`.

We perform static analysis on the 93 SSL-enabled apps and find that 48 of them include `HostnameVerifier` accepting all hostnames. 41 define a verifier that always return true and/or use the `AllowAllHostnameVerifier` subclass, while the other 7 define a hostname verifier that returns true without any check. Our analysis also reveals that 46 of the SSL-enabled apps define a `TrustManager` that actually accepts all certificates. Examples of `TrustManager` and `SocketFactory` doing so are shown in Table 2.

| TrustManager | SocketFactory |
|---|---|
| NaïveTrustManager | NaïveSslSocketFactory |
| BogusTrustManagerFactory | AndroidSSLSocketFactory |
| FakeX509TrustManager | FakeSocketFactory |
| TrustEveryoneTrustManager | TrustNonFacebookSocketFactory |
| TrustAllManager | TrustAllSSLSocketFactory |
| EasyX509TrustManager | EasySSLSocketFactory |
| IgnoreCertTrustManager | SimpleSSLSocketFactory |
| TrivialTrustManager | AllTrustingSSLSocketFactory |
| | BurstlySSLSocketFactory |
| | SelfSignedCertSocketFactory |
| | KakaoSSLSocketFactory |
| | TrustingSSLSocketFactory |
| | ConvivaSSLSocketFactory |
| | SSLSocketFactoryTrustAll |

**Table 2:** Known vulnerable TrustManager and SocketFactory subclasses accepting all certificates [21].

| Sensitive Data | #Apps |
|---|---|
| Username and Password | 3 |
| GPS Location | 4 |
| IMEI Number | 2 |
| IMSI Number | 1 |

**Table 3:** Sensitive data sent via HTTP.

### 5.2 Dynamic Analysis

We start our dynamic analysis by looking for information leakage, i.e., aiming to identify whether any sensitive information is sent in the clear, and then move on to MiTM attacks.

**Unencrypted traffic.** Table 3 summarizes the number of apps sending sensitive information unencrypted. Specifically, we find that4shared, Duolingo, and Pic Collage send usernames and passwords in the clear during login, while Deezer encrypts the password with a nonce and transmits it over HTTP. Location is sent in the clear by MeetMe and Google Maps (no longer the case as of March 2015), GO SMS Pro/Launcher EX, and IMDB. Also, GO SMS Pro sends IMEI and IMSI and Talking Angela the IMEI unencrypted. While it has been reported that transmission of IMEI/IMSI and location data are mostly done by embedded ad libraries [1, 19], this is only true for Talking Angela.

**MiTM.** Our next step is to use Fiddler to mount MiTM attacks. We consider the three scenarios discussed in Section 3, starting with S1, i.e., an attacker having its certificate installed on the user's device. We find that 91 apps establish login connections and give access to secure pages, and leaking sensitive information such as login credentials, financial information, contact list, calendar schedules, and chat messages. On the other hand, 9 apps do not connect thanks to SSL Pinning. In scenario S2 (i.e., an attacker presenting an invalid certificate), 23 of the apps complete the connection (the implementation accepts all certificates), with 9 of them leaking sensitive information. In S3, when the attacker presents a certificate with wrong CN and/or SubjectAltName, we find that 29 of the apps establish a connection, with 11 of them revealing sensitive information. A total of 20 apps are vulnerable in *all three* scenarios, with 9 of these revealing sensitive information.

Only 3 apps (and all in scenario S2) present the user with an error message, such as the one illustrated in Figure 2(a), indicating that the connection was refused due to an SSL certificate error. Other apps keep loading indefinitely, crash, display a message trying to redirect the user to a web browser, display a blank screen or a generic message (e.g., "Unable to connect. Please check your connection and try again", "Incorrect device time. Please ensure the device time is correctly set and try again"). Figure 2(b)–2(d) show examples of unhelpful warning messages prompted by a few apps when the connection is not established.

| Category | #Apps |
|---|---|
| Analyzed Apps | 100 |
| Apps with SSL code | 93 |
| Accepts all certificates | 46 |
| Accepts wrong hostname | 48 |
| Vulnerable to S1 | 91 |
| Vulnerable to S2 | 23 |
| Vulnerable to S3 | 29 |
| Vulnerable to S1, S2, and S3 | 20 |

**Table 4:** Summary of Results.

## 5.3 Summary

The results of our analysis are summarized in Table 4. From the static analysis, we deduce that majority of SSL-enabled apps are vulnerable to MiTM attacks due to wrong hostname verification. Specifically, vulnerable hostname verifiers use the `AllowAllHostnameVerifier` class or return `true` without performing any checks (or checking if the common name ends with a given suffix). We also find that almost half of the apps are vulnerable as `TrustManager` accept invalid or self-signed certificates.

Then, following our dynamic analysis, we posit that: (1) apps with correct implementation of SSL pinning are not vulnerable to MiTM attacks; (2) apps with a vulnerable `TrustManager` establish connections in the presence of an attack; and (3) apps using `AllowAllHostnameVerifier` or with a vulnerable `HostnameVerifier` also establish connections. Also note that the overwhelming majority of apps (91) are vulnerable to powerful adversaries with a certificate on the user's device.

## 6. ANALYSIS

### 6.1 "Secure" Apps

We now analyze the 9 apps that do not establish connections in any of the three attack scenarios considered. These are: Amazon (10M downloads), Barclays Banking (1M), BBM (50M), Bitstrips (10M), Dropbox (100M), MeetMe (10M), TextSecure (500K), Twitter (100M), and Vine (10M), for a total of almost 300M downloads. Note that apps not vulnerable in all attack scenarios account for only 2.6% of the total number of downloads of the apps we test. Also note that since Tweetcaster and Amazon Local/Music are vulnerable in scenario S1, Twitter and Amazon credentials can be compromised even though the Twitter and Amazon apps are not vulnerable.[2]

We also notice that 10 apps (e.g., Skype, Telegram, Viber) employ proprietary protocols or obscure their traffic, and/or did not connect via the MiTM proxy. These apps establish login sessions, and transmit messages/chat conversation in all three attack scenarios, except for Skype which does not load the contact list in scenario S3 (hence, no chat or call conversation can be initiated). Also, the in-app purchase and the WebView interface (when present) of these apps (excluding Telegram, which does not offer in-app purchase) can be exploited in scenario S1. Debit card information used to subscribe and purchase call credit on Skype is also accessible.

### 6.2 Google Apps

Next, we zoom in all Google apps requesting full network access and having 10M+ downloads that are loaded on our Nexus 4 phone by default: Gmail, Calendar, Maps, Google+, Play Store, Play Music, Play Movies (included in our 100-app corpus). We find that they are all vulnerable in scenario S1, with debit card information and PayPal credentials exposed during in-app purchase. Note that this is not limited to the Google apps but also to non-Google apps that use Google's in-app purchase API (a total of 40 in our sample).

---
[2]This is no longer the case for Amazon Local/Music as of March 2015.

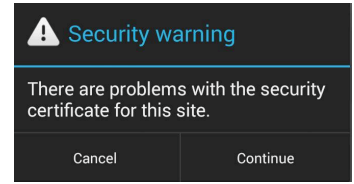

**Figure 3:** Certificate warning displayed by Job Search app in S2 and S3.

Other sensitive information accessible from Google apps in scenario S1 include: usernames and encrypted passwords, email messages (from Gmail), location, calendar, and reminders.

### 6.3 Vulnerable Apps

Looking at apps vulnerable in any of the three attacking scenarios during dynamic analysis, we notice that 59 apps (including Netflix and Facebook[3]) are vulnerable in scenario S1 but not in S2 and S3. This means that these apps are secure against most adversaries, but not against advanced (e.g., state-like) adversaries. Protecting against vulnerability in S1 can be achieved using SSL pinning, which might however add some extra costs.

Also note that there are 32 apps vulnerable in scenario S2 and/or S3, i.e., trusting all certificates or all hostnames, including, e.g., Groupon, Vevo, and Zoosk, that are not vulnerable in S2 but are in S3, leaking user credentials.

Five apps (Booking.com, IMDB, PayPal, Trip Advisor, and TuneIn Radio) establish login sessions in S3, however, we were not able to extract the credentials from the eavesdropped traffic. For instance, after login, PayPal refuses to establish further connections, displaying the message "SSL Error" at the bottom of the screen. We have two possible explanations for this. One is that logins are established simply due to cached cookies. Alternatively, we notice that developers tend to create several different `TrustManager` and `HostnameVerifier` implementing and enforcing different levels of security checks, thus potentially leading to oversights in creating connections with the wrong `TrustManager`.

A total of 20 apps are vulnerable in all three scenarios. One of such app is Jobsearch which displays a warning – shown in Figure 3 – signaling a problem with the certificate and providing users with the option to continue. As highlighted in prior work [36], users tend to ignore SSL warnings, and, if users decide to continue, the adversary would access credentials and other sensitive data.

As mentioned earlier, a few apps send login credentials unencrypted. Specifically, 4shared sends login credentials as part of the URI in the GET command, Deezer encrypts user's password with a nonce before transmission, while Duolingo and Pic Collage send it in plaintext. While developers might not consider their service to be very sensitive, password reuse makes this practice extremely dangerous nonetheless, as discussed in [17].

**Disclosure & Updates.** We communicated our findings to the developers of apps transmitting sensitive information unencrypted and with possible vulnerabilities in scenarios S2 and S3. As a result, Pic Collage now establish login sessions and transmit user information via HTTPS, while 4shared hashes passwords before transmitting them over HTTP. Also, besides updates to Google Maps and Amazon apps discussed earlier, we also report that, as of April 2015, Ask.fm, Textplus and Viber no longer establish connection in scenario S2, while Groupon and IMDB do not in scenario S3.

---
[3]Friend list, chat messages, and credentials are accessible from Facebook in S1.

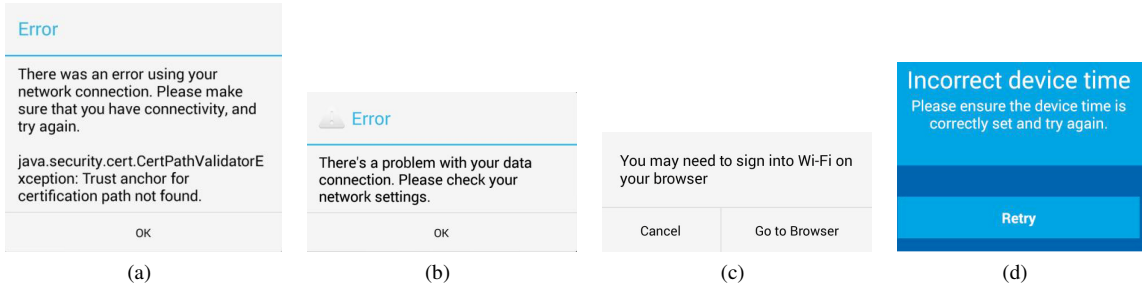

**Figure 2:** Warnings presented to users, as result of non-validation of an SSL certificate.

## 6.4 Static vs Dynamic Analysis

As the results of our static and dynamic analysis differ somewhat, we set to investigate the reasons for this inconsistency.

**Possible false negatives in dynamic analysis.** First, we notice that dynamic analysis may miss some vulnerabilities discovered in static analysis due to vulnerable `TrustManagers` and `SocketFactory` implementations. This happens as vulnerabilities may occur in apps' functionalities that are not tested during our dynamic analysis. For instance, static analysis reveals that `NaïveTrustManager` and `FakeSocketFactory` interfaces are used in 7 apps for the Application Crash Reports for Android (ACRA); however, our dynamic analysis does not test ACRA interaction, thus, it does not report these apps as vulnerable. Similarly, 2 apps use `IgnoreCertTrustManager` for the NativeX mobile ad API.

**Possible false positives in static analysis.** A total of 23 apps possibly contain vulnerable `TrustManager` and `SocketFactory` implementations that dynamic analysis did not report. While 9 are confirmed vulnerabilities, the discrepancy suggests they are not used to create SSL sessions as they might have been introduced for testing purposes but never removed in production. Thus, this might generate some false positives during static analysis, however, because of code obfuscation, it is not always feasible to completely remove such false positives from static analysis.

## 7. DISCUSSION

### 7.1 Self-signed and testing

We now discuss some recommendations to app developers vis-à-vis the vulnerabilities discussed in this paper. We find that many `TrustManagers` are vulnerable as they accept self-signed certificates. This may occur as developers wish to accept self-signed certificates for testing purposes but forget to disable the feature in production. Others purposely choose to employ self-signed certificates in production, and proceed to customize their `TrustManagers`. `TrustManagers` built to accept self-signed certificates usually check for the presence of a single certificate, and verify whether or not it is valid by calling the *checkValidity()* method. This implies that any self-signed certificate currently valid would be accepted by the app.

In order to use self-signed certificates safely, developers should enable SSL pinning instead, rather than using a `TrustManager` that accepts all certificates. While the normal practice of using self-signed certificate is to verify the certificate thumbprint, we recommend the use of self-signed root certificates. Developers should create a `keystore` with self-signed root certificate to sign any number of end-entity certificates to be employed on servers. The `keystore` is then used to create a `TrustManager`. The end certificates are then verified against the self-signed root certificate. Credentials to be checked can include public key, `SubjectAltName`, signature, and any other field specified by the developer (cf. Section 3).

The same method should also be used when developers have a valid certificate, signed by a trusted CA, but need to use a self-signed one for testing purposes. A separate `keystore` containing the self-signed root certificate for development environment should be created and used to initialize the `TrustManager`, but with the same pinning validation logic as that of the production environment. Employing the same pinning validation logic of checking public key, signature, etc., would ensure that developers only add a new self-signed root certificate to the development `keystore` and/or create new end certificates signed by the self-signed root certificate for any number of tests without disrupting the SSL validation logic. If one forgets to change the `keystore` used to initialize the `TrustManager`, the SSL pinning validation logic will act as a fail-safe default, leading to the failure of all secure connections, thus pointing the developer to check the `keystore` used for validation. A sample code snippet that could be used for this purpose is presented in the Appendix.

Finally, whenever a peer's certificate is not known beforehand, developers could use a `TrustManager` that relies on the OS's trusted credentials for the app interface interacting with the peer (unless the peer already provides the API connecting to its servers). This way, using the same `TrustManager` enforcing pinning will result in the HTTPS connection not being established.

### 7.2 Open Problems

Following our analysis, we highlight a few open research problems. First, we emphasize how, more than 1-2 years after prior work drew attention to SSL implementations on mobile platforms, many popular apps still accept all certificates and wrong hostnames, and are vulnerable to MiTM attacks. We argue for the need to give developers more effective tools that can help them detect and fix issues *before* the app is in production, and not only ways to detect vulnerabilities "after the fact." To this end, we have discussed how developers could "safely" use self-signed certificates, and more strategies could be experimented with [12, 22].

The analysis of private information leakage and SSL vulnerabilities should be part of the vetting process performed by app markets, such as Google Play. These already scan apps for malware, inappropriate content, system interference, etc. [28]. Alternatively, research efforts should be encouraged to do so in lieu of app markets. Considering that tools are already available that allow crawling of app markets [37] as well as methodologies for large-scale analysis of SSL vulnerabilities [33], the community could design a portal keeping track over time of such vulnerabilities and reporting them to the public. Ideally, this could also be integrated with results from large-scale dynamic analysis.

Another set of open problems relates to designing meaningful mechanisms for visual feedback. Recall that, in the browser context, the lock icon informs users that their connection is secure and that extensive research has analyzed (and proposed improvements to) SSL warnings [8, 23, 36]. On the contrary, little has been done in the context of smartphone apps. This prompts a number of challenges as it is not clear how to provide meaningful feedback and how to proceed w.r.t. to warnings. Arguably, we should not rely on the user to fix problems the community is not able to, as users have often no idea as to what warnings actually mean or what is the right course of action. A possible research avenue is to contextualize the warnings: if the user is connected via an untrusted Wi-Fi, they could receive a different set of warnings, following a *contextual security* approach [31]. A user that connects to an open Wi-Fi and gets a descriptive warning (e.g., the certificate of www.twitter.com is signed by Mallory, Inc. but was expected to be signed by Verisign) is more likely to disconnect from the Wi-Fi.

Finally, it is well-known that apps often request permissions to access information they do not actually need (as in the notorious Flashlight app case [24]), however, we have usually framed these issue purely in terms of privacy. However, measurement studies could highlight how this can also be exploited by MiTM attackers.

## 8. CONCLUSION

This paper presented a study measuring information leakage from Android apps. Although prior work has highlighted the risks of private information leakage and developed tools for detecting SSL vulnerabilities [13, 33], we found that many of these vulnerabilities are still prevalent, even in popular apps. We analyzed a corpus of 100 Android apps that request full network access and (except for 3 of them) have been downloaded 10M+ times. We decompiled the apps, and analyzed them statically, then, we simulated three MiTM attacks to analyze their security dynamically.

Through static analysis, we found that 46 apps accepted all certificates and 48 failed hostname verification, while, with dynamic analysis, that 91, 23, and 29 apps, respectively, established HTTPS connections with sensitive information leaked in three different attack scenarios. We also found that a few apps established login sessions over HTTP, and that only 3 of the apps provided useful feedback indicating connection failure as a result of an attack.

**Limitations.** While it is inherently hard to perform large-scale dynamic analysis, we acknowledge the limited size of our 100-app corpus. However, this actually represents a reasonably sized sample of popular Android apps – as per statistics in [2], roughly 10% of all apps with at least 10M downloads. Also, we did not use cookies as an attack vector to exploit vulnerabilities: as mentioned in [35], developers include OAuth tokens in cookies and these were often accessible to our MiTM attackers, thus, they could be used to exploit further attack scenarios. Finally, dynamic analysis might produce some false negatives as we might not monitor all possible sockets, while code obfuscation may have masked non-use of vulnerable `TrustManager` or `HostnameVerifier` during static analysis, potentially leading to a handful of false positives.

**Future work.** We plan to include more apps and monitor them over time and are working on a standalone app aiming to counter both passive and active MiTM attacks that extends the features of MITHYSApp [15] and TaintDroid [20]. Specifically, we plan to: (1) block any attempt by apps to transmit tainted information via HTTP, while warning the user, (2) scan the layout of an app's webview to correctly predict and prevent transmission of sensitive information, (3) verify the security of HTTPS connections attempted by apps, and (4) allow transmission of tainted information only after appropriately warning users and getting their consent.

**Acknowledgments.** We wish to thank Jens Krinke, George Danezis, the ACM WiSec anonymous reviewers, and our shepherd Prof. Kevin Butler for their helpful comments, as well as PRESSID for supporting Lucky Onwuzurike.

| | | | |
|---|---|---|---|
| 4shared | eBay | Magic Piano by Smule | Talking Ben |
| 8 Ball Pool | Endomondo Running Cycling Walking | Magisto Video Editor & Maker | Talking Tom 2 |
| Adobe AIR | Facebook | Maps | Tango: Free Video Calls & Text |
| Amazon | Farm Heroes Saga | MeetMe: Chat & Meet New People | The Simpsons: Tapped Out |
| Amazon Local - Deals, Coupons | Flipagram | Facebook Messenger | Telegram |
| Amazon Music | Flipboard: Your News Magazine | Despicable Me (Minion Rush) | Temple Run 2 |
| Talking Angela | Fruit Ninja Free | Netflix | textPlus Free Text + Calls |
| Angry Birds | Gmail | Noom Coach: Weight Loss Plan | TextSecure Private Messenger |
| Ask.fm - Social Q&A Network | GO Launcher EX | PayPal | Google Translate |
| AVG Antivirus Security - FREE | GO Locker - Theme & Wallpaper | PicsArt Photo Studio | TripAdvisor |
| Badoo - Meet New People | GO SMS Pro | Pic Collage | Tumblr |
| BBM | Google+ | Pinterest | TuneIn Radio |
| Barclays Mobile Banking | Google Search | Play Movies & TV | TweetCaster for Twitter |
| Bible | Groupon | Play Music | Twitter |
| Bitstrips | Hangouts | Play Newsstand | VEVO |
| Booking.com Hotel Reservations | Hill Climb Racing | Play Store | Viber |
| Calendar | Instagram | POF Free Dating App | Vine |
| Candy Crush Saga | IMDb Movies & TV | QR Droid Code Scanner | Voice Search |
| Clean Master (Speed Booster) | Jetpack Joyride | Shazam | WhatsApp |
| Deezer Music | Job Search | Skype | WeChat |
| Dictionary.com | KakaoStory | Snapchat | Words with Friends |
| Dropbox | KakaoTalk: Free Calls & Text | SoundCloud - Music & Audio | Yahoo Mail |
| Google Drive | Keep | SoundHound | Yelp |
| Duolingo: Learn Languages Free | LINE: Free Calls & Messages | Spotify Music | YouTube |
| Google Earth | LinkedIn | Subway Surfers | Zoosk |

**Table 5:** Complete list of examined apps.

# APPENDIX

## A. COMPLETE LIST OF APPS

See Table 5.

## B. SSL PINNING USING ANY OF TWO KEYSTORES

```
public final class PinManager implements X509TrustManager {

  /* Get key from keystore */
  KeyStore ks = KeyStore.getInstance(KeyStore.
      getDefaultType());

  /* One of the keystore should be commented... */

  /* Development environment keystore */
  String myStore = "devStore.keystore";
  /* Production environment keystore */
  String myStore = "proStore.keystore";
```

```java
try {
  FileInputStream fis = new FileInputStream(myStore);
      ks.load(fis, password);
      } finally {
      fis.close();
  }

/* Get certificate and associated public key */
Certificate cert = ks.getCertificate(alias);
PublicKey storedPubKey = cert.getPublicKey();
String keyAlgorithm = storedPubKey.getAlgorithm();
private static String pinnedKey = storedPubKey.getEncoded
      ().toString();

/* Create TrustManager with specified keystore */
try {
  TrustManagerFactory tmf = TrustManagerFactory.
        getInstance("X509");
  tmf.init((KeyStore) myStore);
  for(TrustManager tm : tmf.getTrustManagers()) {
  ((X509TrustManager) tm).checkServerTrusted(chain,
       authType);}
}
catch(Exception e) {
  throw new CertificateException(e);
}

public void checkServerTrusted(X509Certificate[ ] chain,
      String authType) throws CertificateException
{
  if(null == chain) {
  throw new IllegalArgumentException("Server
      X509Certificate array is null");
  }

  if(!(null != authType && authType.equalsIgnoreCase(
        keyAlgorithm))) {
  throw new CertificateException("Server AuthType is not
      " + keyAlgorithm);
  }

  /* Get server certificate's public key */
  PublicKey pubKey = chain[0].getPublicKey();
  String receivedKey = pubKey.getEncoded().toString();

  /* Check if keys match */
  final boolean keyMatch = pinnedKey.equalsIgnoreCase(
        receivedKey);
  if(!keyMatch) {
  throw new CertificateException("Public key expected:"+
      pinnedKey+", received: "+receivedKey);
  }
 }
}

/* Initialize SSL context with customized TrustManager */
TrustManager tm [ ] = {new PinManager()};
SSLContext context = SSLContext.getInstance("TLS");
context.init(null, tm, null);

/* Create connection using customized socket factory*/
URL url = new URL("https://www.bob.com/");
HttpsURLConnection connection = (HttpsURLConnection) url.
      openConnection();
connection.setSSLSocketFactory(context.getSocketFactory());
```